\def\po{{p_{\rm{0}}}} 
\def\bpo{{\overline {p_{\rm{0}}}}}
 
\def\bpow{{\overline {p_{\rm{0w}}}}}
\def\exs{{s^{\rm{ex}}}}
\def\rh{{\rho_h}}
\def\dbx{{D_{\rm{bulk}}^x}}

\documentclass[jpcm,12pt]{iopart}
\usepackage{iopams,graphicx}

\begin{document}
\title[Static properties that predict dynamics of confined fluids]{Available states and
  available space: Static properties that predict dynamics of confined fluids}
\author{Gaurav Goel$^1$, William P Krekelberg$^1$, Mark J Pond$^1$,
  Jeetain Mittal$^2$, Vincent
 K Shen$^3$, Jeffrey R Errington$^4$ and  Thomas M Truskett$^{1,5}$}
\address{$^1$ Department of Chemical Engineering, The University of
  Texas at Austin, Austin, Texas 78712}
  \address{$^2$ Laboratory of Chemical Physics, National Institute
  of Diabetes and Digestive and Kidney Diseases,
  National Institutes of Health, Bethesda, Maryland 20892-0520, USA}
  \address{$^3$ Chemical and Biochemical Reference Data Division, NIST, Gaithersburg, MD 20899-8320 USA}
  \address{$^4$ Department of Chemical and Biological Engineering,
  University at Buffalo, The State University of New York, Buffalo,
  New York 14260}
  \address{$^5$ Institute for
  Theoretical Chemistry, The University of Texas at Austin, Austin,
  Texas 78712}
\ead{truskett@che.utexas.edu}




\begin{abstract}
  Although classical 
density functional theory provides reliable predictions for
  the static properties of simple equilibrium fluids under
  confinement, a theory of comparative accuracy for the transport
  coefficients has yet to emerge.  Nonetheless, there is evidence that
  knowledge of how confinement modifies static behavior can aid in
  forecasting dynamics.  Specifically, recent molecular simulation
  studies have shown that the relationship between excess entropy and
  self diffusivity of a bulk equilibrium fluid changes only modestly
  when the fluid is isothermally confined, indicating that knowledge
  of the former might allow semi-quantitative predictions of the
  latter.  Do other static measures, such as those that characterize
  free or available volume, also strongly correlate with
  single-particle dynamics of confined fluids?  Here, we investigate
  this question for both the single-component hard-sphere fluid and
  hard-sphere mixtures.  Specifically, we use molecular simulations
  and fundamental measure theory to study these systems at
  approximately $10^3$ equilibrium state points.  We examine three
  different confining geometries (slit pore, square channel, and
  cylindrical pore) and the effects of particle packing fraction 
and particle-boundary interactions.
  Although average density fails to predict some key qualitative
  trends for the dynamics of confined fluids, 
 we find that a new generalized measure of
  available volume for inhomogeneous fluids strongly correlates with 
the self diffusivity across
  a wide parameter space in these systems, approximately independent
  of the degree of confinement.  
An important consequence,
  which we demonstrate here, is that density functional theory
  predictions of this static property can be used together with knowledge of bulk fluid
  behavior to semi-quantitatively estimate the diffusion coefficient of
  confined fluids under equilibrium conditions.  
\end{abstract}
\maketitle

\section{Introduction}
\label{sec:intro}

Confining an equilibrium fluid of particles to length scales on the
order of several particle diameters changes both its static and
dynamic properties.  Classical density functional theory (DFT) can
often make reliable predictions concerning the former, but
implications of confinement for dynamics remain challenging to
forecast for even the most basic models.  For example, Enskog theory
generalized for inhomogeneous fluids~\cite{Davis1987Kinetictheoryof}
predicts that constant-activity confinement of the hard-sphere (HS)
fluid between hard walls will significantly decrease self-diffusivity
parallel to the
boundaries~\cite{Vanderlick1987Self-diffusioninfluids}. Recent
molecular dynamics simulations of that system, however, show that this prediction is
qualitatively incorrect, i.e., self diffusivity increases with
constant-activity confinement~\cite{Mittal2007Doesconfininghard-sphere}.  Hydrodynamic
theories, on the other hand, can predict how the presence of a single
wall~\cite{Happel1973LowReynoldsNumber} or confinement between two
walls~\cite{Faucheux1994ConfinedBrownianmotion,Lin2000Directmeasurementsof,Dufresne2001Browniandynamicsof,Saugey2005Diffusioninpores}
impacts the self diffusion of a single Brownian particle in solvent.
But it is not yet clear how to generalize these approaches to account
for the effects that strongly inhomogeneous static structuring has on
the transport coefficients of dense confined
fluids~\cite{Mittal2007Doesconfininghard-sphere,Goel2008Tuningdensityprofiles,Mittal2008LayeringandPosition}.
Given the absence of a reliable microscopic theory, the development 
of
new qualitative or semi-quantitative heuristics for predicting
dynamics of confined fluids would be of considerable practical use.

In this spirit, one productive strategy has been to first exhaustively
simulate the static and dynamic behaviors of simple models of confined
fluids.  The idea is that comprehensive study of these systems may reveal 
static quantities that strongly correlate with transport
coefficients for a wide variety of confining environments.  Knowledge
of these correlations together with reliable predictions of the static
quantities from equilibrium theory would, in turn, lead to indirect predictions
for how confinement modifies dynamics.

Indeed, recent simulation data covering hundreds of state points for
various confined HS, Lennard-Jones, and square-well fluids
point to the existence of a robust, isothermal correlation between the self-diffusion
coefficient $D$ and the excess entropy per particle (over ideal gas) $s^{ex}$ 
--- a relationship that is approximately independent of the degree of
confinement for a wide range of equilibrium conditions~\cite{Mittal2007Doesconfininghard-sphere,
  Goel2008Tuningdensityprofiles, Mittal2006ThermodynamicsPredictsHow,
  Mittal2007RelationshipsbetweenSelf-Diffusivity, Mittal2007Confinemententropyand}.
The practical implication is that the mathematical form of the
correlation for a given fluid (obtained from bulk fluid simulations) can
be used together with knowledge of the excess entropy of the confined
system (computed, e.g., via DFT) to make
semi-quantitative predictions for how confinement will modify the
self diffusivity.
As has been discussed
elsewhere~\cite{Mittal2007Doesconfininghard-sphere,Goel2008Tuningdensityprofiles}, 
this strategy can
successfully predict subtle, confinement induced effects of packing
frustration on relaxation processes, behavior not reflected in other static
quantities traditionally thought to track dynamics, such as the
average density.


Although a fundamental and general derivation that explains the observed
relationship between excess entropy and dynamics
of confined fluids is still lacking, the fact that the two are
connected is not surprising.  Excess entropy
is a negative quantity that measures the extent to which static
interparticle correlations reduce the number of microstates available
to the fluid.  In fact, $-\exs$ is often used as a metric for
characterizing the ``amount'' of structural order present in condensed
phase
systems~\cite{Truskett2000Towardsquantificationof,Mittal2006QuantitativeLinkbetween,Errington2006Excess-entropy-basedanomalieswaterlike,Krekelberg2007Howshort-rangeattractions,Lipkowitz2007ReviewsinComputational}.
Since interparticle correlations strongly influence collisional
processes, it makes intuitive sense that macrostate changes
which increase structural order ($-\exs$) might also tend to reduce
single-particle mobility.  Qualitative arguments such as these have
previously been presented to rationalize empirically observed 
correlations between excess entropy and transport coefficients of 
both bulk~\cite{Rosenfeld1977Relationbetweentransport,Rosenfeld1999quasi-universalscalinglaw,
   Dzugutov1996universalscalinglaw} and confined~\cite{Mittal2007Doesconfininghard-sphere,Goel2008Tuningdensityprofiles, Mittal2008LayeringandPosition,Mittal2006ThermodynamicsPredictsHow,Mittal2007RelationshipsbetweenSelf-Diffusivity,Mittal2007Confinemententropyand}
equilibrium fluids. Nonetheless,  it is natural to wonder whether excess
entropy is unique in this regard.  Perhaps other measures also
accurately forecast the implications of confinement for the dynamics
of dense fluids.  For example, does the mobility of inhomogeneous HS
fluids also increase with the average amount of space available for
particle motion?  Do some measures of free or available volume
correlate much more strongly with dynamics than others?  Can one
quantitatively, or semi-quantitatively, predict self diffusivity of
confined fluids based on knowledge of how confinement affects the free
or available volume of the system?

Here, we put the above questions to stringent tests for a variety of
confined fluid systems. To ascertain the effect of pore geometry on
correlations between dynamics and thermodynamics, we study a
monodisperse hard-sphere fluid confined to smooth hard-wall slit
pores, square channels, and cylindrical pores. We also explore the
effects of boundary interactions in the slit-pore geometry by
examining cases for which square-well (attractive) or
square-shoulder (repulsive) particle-boundary interactions are present.  Finally, we investigate the
dynamics and thermodynamics of a confined binary hard-sphere mixture
which can be equilibrated in the fluid state at higher packing
fractions without
crystallizing than the corresponding single-component fluid.
Together, this study represents, to our knowledge, the most
comprehensive exploration of the relationships between static and
dynamic properties in these basic inhomogeneous systems to 
date.  We calculate
the self-diffusion coefficient [via molecular dynamics simulations],
and excess entropy and various measures of available volume [via
Transition Matrix Monte Carlo (TMMC) simulations and fundamental
measure theory] at approximately
$10^3$ state points.  Our results illustrate that predictions of
either excess entropy or 
a generalized measure of average available volume from classical 
density functional theory can be used together with knowledge of bulk fluid
behavior to semi-quantitatively predict the diffusion coefficient of
confined fluids across a wide range of equilibrium conditions.

\section{Connections between density, available volume, and dynamics:
  Preliminary evidence}
\label{connection_preliminary}
The available volume in a configuration of the single-component 
HS fluid refers to the
space into which the center
of an additional HS particle of equal size can be inserted without
creating an overlap with existing particles or the boundary.  
It might also be considered a coarse measure of the space immediately 
available for particle motion in that configuration of the system.  
For the bulk equilibrium HS fluid, the ensemble-averaged 
fraction of the total volume 
that is available (in the above sense) is given by $p_0=\rho/\xi$~\cite{Widom1963SomeTopicsin}, where $\rho$ is number density,
$\xi=\exp[\beta\mu]/\Lambda^3$ is activity, $\mu$ is the chemical
potential, $\beta=[k_{\mathrm{B}}T]^{-1}$, $k_{\mathrm{B}}$ is
the Boltzmann constant, $T$ is the temperature, and $\Lambda$ is the thermal deBroglie wavelength.
It is known that increasing $\rho$ monotonically decreases both $p_0$
and 
self diffusivity $D$ 
for this system; i.e. $d p_0/ d \rho<0$ and $d D / d \rho < 0$ (see,
e.g. \cite{Lipkowitz2007ReviewsinComputational}), which is consistent
with the intuitive idea that these static and dynamic properties
might be linked.  

\begin{figure}[h]
  \includegraphics{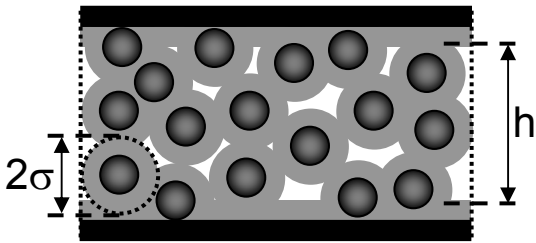}
  \caption{Schematic of an inhomogeneous HS fluid confined between
    boundaries in a slit pore geometry.  
    The walls are placed a distance $H=h+\sigma$ apart,  
    where $h$ is the length accessible to particle centers.  
    Dark regions indicate hard spheres and
    confining walls.  Additional particle centers are excluded from
    the grey (overlap) region.
    The white region indicates the volume available for inserting an
    identical hard sphere of diameter~$\sigma$. 
    \label{freevolschematic}}
\end{figure}

Are density, available space, and self diffusion connected in the same
qualitative way for inhomogeneous HS fluids?  Previous studies have
provided some information useful for addressing this question.  For
example, it is known that confining the equilibrium HS fluid between
hard walls (while maintaining fixed $\xi$) significantly increases
the average particle density, i.e. $(\partial \rh /
\partial h^{-1})_{\xi}>0$, over a wide range of
$\rh$ and $h$~\cite{Vanderlick1987Self-diffusioninfluids,Mittal2007Doesconfininghard-sphere}. Here, $\rh=N/(Ah)$ is average density, $N$ is the total number of particles,
$A$ is the area of the interface between the fluid and one hard wall,
and $h$ is the center-accessible width of the slit pore [i.e., not
including the ``dead space'' that the particle centers are 
excluded from due to their interaction with the boundaries (see
Figure~\ref{freevolschematic})].  The fact that $\rh$ increases 
upon constant-$\xi$
confinement initially appeared consistent
with earlier kinetic theory predictions that $D$ of this system would
show a corresponding 
decrease~\cite{Vanderlick1987Self-diffusioninfluids}. However,
recent simulation data have demonstrated that the kinetic theory predictions
were qualitatively 
incorrect~\cite{Mittal2007Doesconfininghard-sphere}.  That is,
both $\rh$ and $D$ typically increase upon constant-$\xi$ confinement
[$(\partial D/\partial \rh)_{\xi}>0$, the fluid
gets ``denser'' on average and particles diffuse {\em more rapidly}],
a trend that is the opposite of what might be expected based on the
bulk HS fluid behavior.  Interestingly, confined HS fluids show a
different trend when an alternative thermodynamic 
constraint
is applied.  Specifically, if $h$ is held fixed, then increasing $\rh$ has the
effect of {\em decreasing} $D$, i.e., $(\partial D/\partial \rh)_{h}<0$]~\cite{Mittal2007Doesconfininghard-sphere,Mittal2006ThermodynamicsPredictsHow}.
This preliminary data illustrates that knowledge of how $\rh$ changes
is not, in and of itself, 
enough to even qualitatively predict the implications of
confinement for the dynamics of a fluid.  Indeed, in
Section~\ref{sec:diffpredictions} of this article, we present
extensive numerical evidence for a variety of confined fluid systems
which underscores this point.  We also explore whether adopting a
definition of average density based on the total rather than 
center-accessible volume of the pore (see
also~\cite{Mittal2007Doesconfininghard-sphere,Mittal2006ThermodynamicsPredictsHow})
improves predictions for how confinement modifies dynamics.

Does available volume show a more reliable correlation to dynamics than
average density?  
The fractional available volume in an inhomogeneous fluid is 
inherently a local quantity,  and it is given by $\po _i
(z)=\rho _i(z) \exp[\beta u^w_\mathrm{i} (z)]/\xi _i$, 
where $\rho _i(z)$ and
$u^w_\mathrm{i} (z)$ represent the singlet (one-particle) density and
the wall-particle interaction potential for species $i$, 
respectively, evaluated  at a distance $z$ from one
wall~\cite{Lebowitz1965ScaledParticleTheory,Widom1978Structureofinterfaces,Hendersen1983StatisticalMechanicsOf}.
The volume averaged quantity can be expressed as
\begin{equation}
  \label{eq:po_hshw}
\bpo_i\equiv \frac{1}{V_\mathrm{c,i}}\int_{V_\mathrm{c,i}} \po_i dV
\end{equation}
where the integral is over the particle-center-accessible volume
$V_\mathrm{c,i}$.  For the special case of a single-component 
HS fluid confined 
between  smooth hard walls, Eq.~\eref{eq:po_hshw} 
reduces to $\bpo=\rh/\xi$.  Note that since $ \left(\partial \bpo / \partial h^{-1}\right)_\xi=\xi^{-1} \left(\partial \rh / \partial
    h^{-1}\right)_\xi$, and  $\left(\partial \rh / \partial
    h^{-1}\right)_\xi>0$ across a wide range of $\rh$ and $h$~\cite{Vanderlick1987Self-diffusioninfluids,Mittal2007Doesconfininghard-sphere}, it follows that $\left(\partial
    \bpo/\partial h ^{-1} \right) _\xi>0$ for those conditions.  This
  increase in the fraction of available space with constant-$\xi$
  confinement provides a simple physical
  explanation for the counterintuitive observation that 
$D$ {\em increases} along the same thermodynamic path.
  The density-dependent behavior of the $\bpo$ under the constraint of
  constant $h$ is also qualitatively consistent with the dynamical
  trends of the confined HS fluid.  In particular, both $\left(\partial \bpo/\partial
    \rh \right)_h<0$ and $\left(\partial D/\partial
    \rh \right)_h<0$.  All of this strongly suggests that 
$\bpo$ is a more relevant 
static metric for particle mobility than the average particle 
density $\rh$.

In the following
sections, we test the generality of these preliminary observations
by carrying out an extensive quantitative comparison of 
the correlations between self diffusivity $D$ and
various static measures (density, excess entropy, and fractional
available space) for single-component and binary HS fluids confined to
a variety of channels with different geometries and particle-boundary 
interactions.  The results clarify which of these static quantities
reliably predict the implications of confinement for
single-particle dynamics.  


\section{Methods}
\label{sec:methods}

We study single-component HS fluids of particles with diameter $\sigma$
both in the bulk and confined to channels with three types of geometries:
(i) quasi-two-dimensional slit pores, (ii) quasi-one-dimensional
square channels, and (iii) cylindrical pores.  Specifically, we
consider (i) seven slit pores with thickness $H/\sigma=5, 6, 7, 8, 9,
10,~\mathrm{and}~15$ in the confining $z$ direction (see 
Figure~\ref{freevolschematic}) together with periodic boundary conditions in 
the $x$ and $y$ directions, (ii) seven 
square channels with total $x-y$ cross-sectional dimensions of $H^2/\sigma^2
=25, 36, 49, 64, 81, 100,~\mathrm{and}~225$ together with a 
periodic boundary condition in the  
$z$ direction, and (iii) six cylindrical channels of total diameter
$H/\sigma=6, 7, 8, 9,
10,~\mathrm{and}~15$ together with a periodic boundary condition 
in the axial $z$ direction. 

We take the interaction of a particle with a channel wall $u_\mathrm{w} (s)$
in all cases to have a generic square-well form:
\begin{equation}
  \label{sw_interaction}
  u_\mathrm{SW}(s) = \cases{\infty& $s < \sigma /2$\\
    \epsilon_\mathrm{w}& $\sigma/2 \le s < \sigma$\\
    0& $s \ge \sigma$\\}
\end{equation}
where $s$ is the
shortest distance between the particle center and the wall of
interest.  For all three geometries, we study the case 
$\epsilon_\mathrm{w}=0$, i.e. smooth hard boundaries. Additionally, for the
slit pore with
size $H/\sigma=5$, we investigate cases with $\epsilon_\mathrm{w}=2
k_{\mathrm B}T$ (a repulsive shoulder) and
$\epsilon_\mathrm{w}=-2k_{\mathrm B}T$ (an attractive well).

We also consider a binary HS mixture confined between smooth hard
walls in a slit pore of width $H/\sigma_1=5$. 
For this system, the particle diameter ratio is taken to be $\sigma_\mathrm{2}/\sigma_\mathrm{1}=1.3$ and the particle masses are 
proportional to their volume, 
i.e., $m_\mathrm{2}/m_\mathrm{1}=(\sigma_\mathrm{2}/\sigma_\mathrm{1})^3$.
These parameter values closely mimic those examined in 
recent experiments of
binary colloidal mixtures under
confinement~\cite{Nugent2007ColloidalGlassTransition}.   


To explore dynamic properties of these fluids, we perform
molecular dynamics simulations using an event-driven
algorithm~\cite{Rapaport2004TheArtOf} in the microcanonical ensemble
with $N=4000$ particles for monodisperse HS systems and $N=3200$ for
the binary HS systems.  For bulk systems, we use a cubic
simulation cell of volume $V$. For the confined systems, we adopt a
rectangular parallelepiped simulation cell of volume $V=H_x H_y H_z$
with appropriate boundary conditions depending on geometry.  
We extract the self diffusivity $D$ by
fitting the long-time ($t >> \sigma_\mathrm{1}
\sqrt{m_\mathrm{1}\beta}$) mean-squared displacement to the
Einstein relation $\big<\Delta
\mathbf{r}^2_{d_p}\big>=2 d_p Dt$, where $\big< \Delta
\mathbf{r}^2_{d_p}\big>$ corresponds to motions in the $d_p$
periodic directions. For the sake of clarity, we reserve the symbol
$D$ for the self diffusivity of fluids under confinement and
$D_{\rm{bulk}}$ for the self diffusivity of the bulk
fluid.  To obtain reliable estimates, we
average self diffusivities over four
independent trajectories. For simplification, we report
quantities from this point forward 
implicitly non-dimensionalized by appropriate combinations
of the characteristic length scale, taken to be the HS diameter of the smallest
particle in the fluid $\sigma_\mathrm{1}$, 
and a characteristic time scale, given by
$\sigma_\mathrm{1} \sqrt{m_\mathrm{1}\beta}$.  Thus, all energies are
implicitly per unit $k_{\mathrm B} T$, and $T$ is effectively scaled out of
the problem altogether.

We obtain thermodynamic properties using grand-canonical
transition-matrix Monte Carlo (GC-TMMC) simulation.  For pure fluids,
we use an algorithm presented by
Errington~\cite{Errington2003DirectCalculationOf} and for binary 
mixtures we employ a strategy developed by Shen and Errington~\cite{Shen2005Determinationoffluid},
wherein one combines a series of semigrand ensemble simulations to
construct the system's free energy over a wide range of densities and
compositions.  We conduct GC-TMMC simulations in a standard grand
canonical ensemble where the volume $V$, temperature $T$, and 
activities \{$\xi _1,\xi _2$\} are held constant and the particle numbers
\{$N _1, N _2$\} and energy $E$ fluctuate. For notational convenience,
we denote the sets \{$N _1, N _2$\} and \{$\xi _1,\xi _2$\} as
$\mathbf{N}$ and $\boldsymbol{\xi}$, respectively, using conventional
vector notation. The activity of component $i$ is
defined as $\xi _i = \Lambda _i ^{-3}\exp(\mu _i)$, where $\mu _i$
is the chemical potential and $\Lambda _i$ is the thermal de Broglie
wavelength.  For the pure component GC-TMMC simulations 
we present here,
we set $\xi=1$ and adjust the particle center-accessible volume
$V _\mathrm{c}$ to make the total volume $V \approx 1000$.
Simulations of the binary mixture use $V=125$ and $V=245$ for the 
bulk and confined fluids, respectively. 
For the bulk and confined simulations, we set $\xi _1=173.7$ 
and $\xi _2=381.5$. 

The key quantity we extract from the GC-TMMC simulations is the
particle number probability distribution ${\Pi} (\mathbf{N})$. 
Once we obtain this distribution, we use basic
statistical mechanics principles and histogram
reweighting~\cite{Ferrenberg1988NewMonteCarlo} to evaluate
thermophysical properties over a range of activity values.  
First, we use histogram reweighting to deduce 
${\Pi} (\mathbf{N})$ at a set of activities 
$\boldsymbol{\xi}_\mathrm{new}$ generally different from that of the
GC-TMMC simulation $\boldsymbol{\xi}_\mathrm{sim}$,
\begin{equation}
  \label{eq:PiN}
  \ln {\Pi} (\mathbf{N};\boldsymbol{\xi}_\mathrm{new}) = 
  \ln {\Pi} (\mathbf{N};\boldsymbol{ \xi}_\mathrm{sim}) +
  \sum_i N_i (\ln \xi _{i, \mathrm{new}} - \ln \xi _{i,
      \mathrm{sim}}),
\end{equation}
where it is understood that the probability distributions are not
normalized.  We obtain average particle numbers
$\left<\mathbf{N}\right>$ from first-order moments of ${\Pi} (\mathbf{N})$,
\begin{equation}
  \label{eq:avgN}
  \left<\mathbf{N (\boldsymbol{\xi})}\right> = \sum\mathbf{N}{\Pi}(\mathbf{N;
    \boldsymbol{\xi}}) / \sum  {\Pi}(\mathbf{N; \boldsymbol{\xi}}).
\end{equation}
We calculate $\mathbf{\rho}$ and $\mathbf{\rho} _\mathrm{h}$ via
normalization of $\left<\mathbf{N}\right>$ by $V$ and $V _\mathrm{c}$,
respectively.

We define excess entropy as the difference
between the fluid's entropy and that of an ideal gas with the same
density profile.  The particle number
probability distribution provides the density and composition
dependence of the Helmholtz free energy at a given temperature.
Therefore, we combine knowledge of ${\Pi}(\mathbf{N})$, average excess
configurational energies $U^{\rm{ex}} (\mathbf{N})$, and
particle-number-specific spatial density distributions $\rho
(\mathbf{N,r})$ to obtain the total excess entropy
$S^{\mathrm{ex}}$~\cite{Mittal2006ThermodynamicsPredictsHow,Mittal2007Confinemententropyand,Errington2006Excess-entropy-basedanomalieswaterlike},
\begin{eqnarray}
  \fl S^{\mathrm{ex}} (\mathbf{N}) = U^{\rm{ex}}(\mathbf{N}) + \ln {\Pi}
  (\mathbf{N})/{\Pi} (\mathbf{0}) \nonumber\\
  + \sum_i \left\{\ln N_i! - N_i \ln \xi_i -
  N_i \ln N_i + \int \rho_i (\mathbf{N,r})\ln\rho_i
  (\mathbf{N,r})d\mathbf{r}\right\}\label{eq:TMMCexS}
\end{eqnarray}

We also predict the thermodynamic quantities of the bulk
single-component and binary HS fluids using the
Carnahan-Starling~\cite{carnahan:635} and
Boublik-Mansoori-Carnahan-Starling-Leland~\cite{boublik:471,mansoori:1523}
equations of state, respectively.  We predict the thermodynamic 
properties of confined HS fluids
using a recent modification~\cite{Yu2002Structuresofhard-sphere} of
Rosenfeld's fundamental measure
theory~\cite{Rosenfeld1989Free-energymodelinhomogeneous}. Fundamental
measure theory is a classical DFT of inhomogeneous fluids that
has been shown to accurately predict structure and thermodynamics of
confined HS systems in various restrictive geometries up to very high
densities~\cite{Gonzalez2006Densityfunctionaltheory}. For numerical
evaluation of the DFT for slit and
cylindrical pores, we use Picard iterations on a grid spacing 
of  $0.005$.  We update densities according to, 
$[\rho]_{n+1}^{\mathrm{in}}=0.95[\rho]_n^{\mathrm{in}}+0.05[\rho]_n^{\mathrm{out}}$, where
$[\rho]_{n+1}^{\mathrm{in}}~\mathrm{and}~[\rho]_n^{\mathrm{in}}$ are the input density
profiles at the $n+1^{th}$ and $n^{th}$ iterations, respectively, 
and $[\rho]_n^{\mathrm{out}}$ is the output density profile at 
the $n^{th}$ iteration. We stop Picard iterations when the
relative change in output density profile between two successive
cycles [1 cycle = 20 iterations] becomes less than $10^{-5}$. For
numerical evaluation of the DFT
in the square channel geometry, we use Sandia National Laboratories 
Tramonto package~\cite{Sears2003ANewEfficient}. 
We adopt a grid of $0.05 \times 0.05$ for $H=5,6$ and a
grid of $0.075 \times 0.075$ for $H>6$. 
We stop the minimization algorithm when the relative 
or absolute change in the grand potential is less than $10^{-7}$.

 \section{Testing structure-property relations for predicting
   self diffusivity of confined fluids}
 \label{sec:diffpredictions}


In this section, we explore the accuracy of the following strategy 
for predicting the self diffusivity of confined HS fluids:  (1) 
determine the
value of a static quantity~$x$ of a confined fluid believed to be
relevant for dynamics (e.g., its density,
excess entropy, or fractional available volume), and (2) input this
value into the 
relationship between self diffusivity and that same static 
quantity for the bulk
fluid, $\dbx (x)$, 
to estimate the confined fluid self diffusivity, $D$.
Of course, such a strategy can only provide approximate predictions.
While there is a one-to-one relationship for the equilibrium HS fluid 
between the 
self diffusivity
and any one of the 
aforementioned static quantities,{\footnote{This is true if the self diffusivity is 
appropriately non-dimensionalized, as we have done here, 
to remove the trivial effect of the thermal velocity of the
particles.}
the dynamic properties of the confined fluid generally depend 
on a larger number of 
variables (e.g., the dimensions of the confining geometry,
the nature of the particle-boundary interactions, and 
the chemical potential).  Nonetheless, the hope is that one can
discover a static quantity $x$ whose relationship with $D$ is 
largely insensitive to the effect of confinement.  If so, the
bulk structure-property relationship $\dbx$ 
can be used to predict $D$
independent of the other parameters of the confined 
system.  Systematic tests of this idea should give 
new insights into the structural properties 
that are most relevant for single-particle dynamics of inhomogeneous fluids.

To investigate the accuracy of predictions by 
this approach, we 
use the ``exact'' results of 
molecular simulations to examine the ratio of the bulk fluid
self diffusivity to that of confined fluids with the same value of
$x$, i.e., $\dbx/D$.  Since one is often
interested in both $D$ and the effective characteristic time associated with
diffusive motion ($D^{-1}$), we present plots of $\dbx/D$ in this work on a 
logarithmic scale, a representation for which 
overpredictions and underpredictions of $D$ by
the same factor appear the same distance from unity.  We also 
present statistics associated with the relative errors of the 
predictions for different $x$.  
For each $x$ that we consider here, 
we analyze roughly $10^3$ state points
of the equilibrium HS fluid confined to the various pore geometries
described in Section~\ref{sec:methods}.  This
data, when taken as whole, spans approximately four decades in $D$.

\subsection{Confinement in channels with
  smooth hard boundaries}
\label{sec:dbydbulk_neutral}

\begin{figure}[h]
  \includegraphics{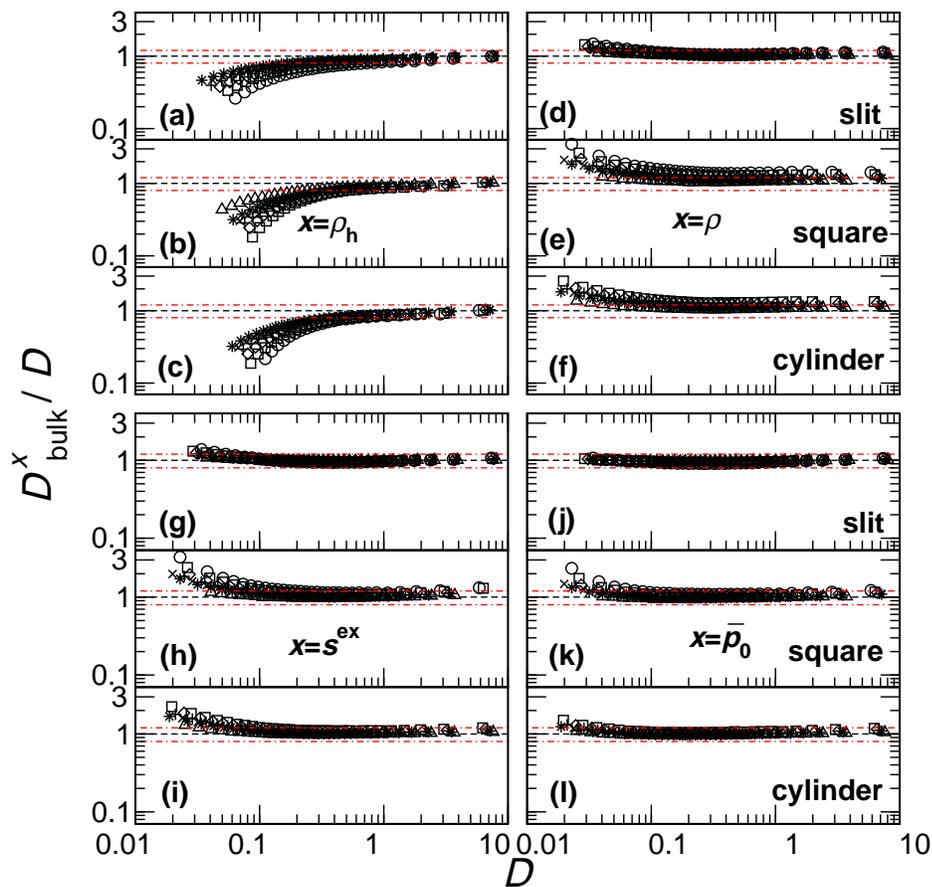}
  \caption{Ratio of self diffusivity of a bulk HS fluid to that of a
    confined HS fluid with the same value of a static quantity $x$, 
    $\dbx/D$, obtained via molecular 
    simulations. Data are shown for the fluid confined to 
    slit pore [a, d, g, j],  square channel [b, e, h, k], and 
    cylindrical pore [c, f, i, l] geometries. 
The static quantity $x$ is indicated in each of the four panels. 
    $20\%$ bounds on relative error in confined fluid 
self diffusivity ``predictions'' by using
    the bulk structure-property relation $\dbx$ 
    are shown by red dash-dotted lines.
    The equilibrium fluid states shown here span the density range 
    $0<\rh<\rho_0$, where $\rho_0 \approx 0.95$ for (a-c) and
    $1<\rho_0<1.25$ for (d-l), depending on pore size.  Pores shown have
    confining dimensions of $5$ (\opencircle), $6$ (\opensquare), $7$
    (\opendiamond), $8$ (+), $9$ (x), $10$ (*), and $15$
    (\opentriangle).  These dimensions correspond to channel width $H$
    for slit pores and square channels and channel diameter $d$ for 
    cylindrical pores.  All channels have smooth hard boundaries.
    \label{dbydbulk_all}}
\end{figure}

Here, we examine the ratio $\dbx/D$ for the single-component HS fluid confined to various
geometries by smooth hard boundaries [i.e. $\epsilon _{\mathrm{w}}=0$, see \eref{sw_interaction}].  
We begin by testing the predictions that follow from
assuming that $x=\rh$, the 
number density averaged over the particle-center-accessible volume of the
pore, is the relevant static metric for dynamics [see
Figure~\ref{dbydbulk_all}(a-c)].   It is immediately clear from the
data that
$\rh$ does not, in itself, provide a good basis for prediction.  HS fluids 
confined in slit-pore, square-channel, and cylindrical 
geometries generally
exhibit a wide range of $D$ for each $\rh$, with the fastest dynamics
occurring in the smallest pores.  In fact, note that 
the bulk structure-property relation $D_{\mathrm{bulk}}^{\rh}$
can underpredict $D$ by nearly a factor of ten for fluids in the most 
restrictive geometries.  The performance of the bulk structure-property 
relation using $x=\rh$ 
is actually even worse than it appears in
Figure~\ref{dbydbulk_all}(a-c) for the following reason.  
The freezing transition occurs at a 
density of $0.945$ for the bulk
HS fluid, which provides an upper limit on values of $\rh$ that 
can be used for predictions using $D_{\mathrm{bulk}}^{\rh}$.  However, center-accessible 
densities for the equilibrium fluid 
in the smallest square-channel and cylindrical pores can reach as high
as $\rh \approx 1.25$.  Thus, the bulk structure-property relation
$D_{\mathrm{bulk}}^{\rh}$ {\em cannot even make predictions} 
for a significant fraction of the equilibrium
state points for highly confined HS fluids.  
All of this confirms the
preliminary expectation discussed in
Section~\ref{connection_preliminary} that knowledge of $\rh$ and bulk
fluid behavior is not enough to predict the self diffusivity 
of confined fluids. This conclusion is consistent with the
earlier observations of 
Mittal et al.~\cite{Mittal2006ThermodynamicsPredictsHow} concerning
a smaller set of data for the HS fluid confined to slit pores.  

In Figure~\ref{dbydbulk_all} (d,g, and, j), we again present 
$\dbx/D$ for the HS fluid confined to slit pores, 
but now $\dbx$ is the corresponding bulk fluid relation between
diffusivity and one of three alternative static properties ($x$):  
average density based on total pore volume $\rho=\rh (1-H^{-1})$,
excess entropy per particle $\exs$, and fraction of available volume 
$\bpo$.  The data in these plots corresponds to confined fluid states with 
packing fractions that vary from the dilute gas to the 
freezing transition for pore widths $H \ge 5$.  As can be seen, each
of these static measures can
provide semi-quantitative predictions for confined fluid diffusivities
when used together with the corresponding bulk structure-property correlation.
In fact, for $93\%$~($x=\rho$), $97\%$~($x=\exs$), and
$100\%$~($x=\bpo$) of equilibrium state points for these systems,  
the predictions provided by $\dbx$ are within $20\%$ of
the ``exact'' MD data for $D$.   Note that the 
very small fraction of overpredictions based on $\rho$ or $\exs$ that
exceed $20\%$ relative error correspond 
to the high density, low $D$ state points near the freezing
transition.

Based on the slit-pore data, it might be tempting to conclude that 
total-volume-based average density~$\rho$ tracks dynamics nearly as
reliably as~$\exs$ and 
$\bpo$ for confined fluids.  To provide a more stringent test of 
this preliminary conclusion, we now examine $\dbx/D$
for HS fluids confined to quasi-one-dimensional square channel and cylindrical pore geometries with
edge dimensions $H \ge 5$ and diameters $d \ge 6$, respectively.  
Fluids confined in these geometries
have a significantly higher percentage of particles near the
boundaries than in the corresponding slit pores, and hence the effects of confinement on both structure and
dynamics should be more 
pronounced.  

Figure~\ref{dbydbulk_all}(e,f) shows that for square channel and
cylindrical geometries, self-diffusivity predictions
based on $D_\mathrm{bulk}^{\rho}$ can be significantly higher
than the actual $D$ of a confined fluid with the same $\rho$.  
In fact, for the densest fluid systems studied here, 
the bulk structure-property relation $D_{\mathrm{bulk}}^{\rho}$ is
between $2$ and $4$ times larger than $D$, depending on $H$.  
Furthermore, Figure~\ref{dbydbulk_all}(h,i,k,l) illustrates that 
$\dbx$ predictions for $x=\exs$ or $x=\bpo$ are in the 
semi-quantitative range for
a larger fraction of state points than those based on $x=\rho$.
Specifically,  $\dbx$ is within  $20\%$ of $D$ for 
$46\%$  ($x=\rho$), $82\%$ ($x=\exs$), and $95\%$ ($x=\bpo$)
of the state points.
The main differences occur for high density, low $D$ state points,
where predictions based on fractional available volume are significantly more 
accurate than those based on excess entropy or density.

\begin{figure}[h]
  \includegraphics{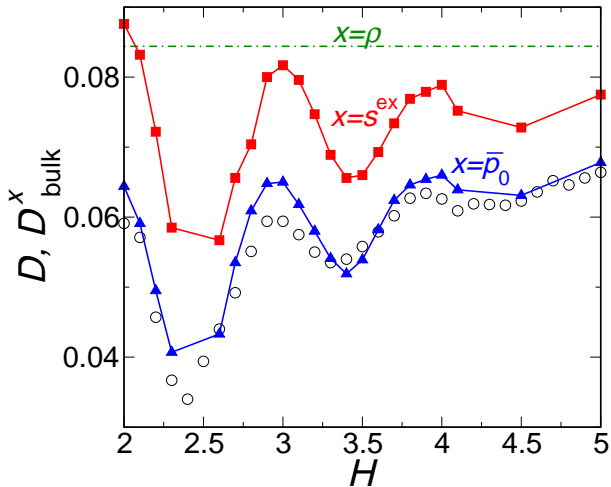}
    \caption{Self diffusivity $D$ 
  of a HS fluid confined in narrow slit pores of
  width $H=2-5$ by smooth hard boundaries. The density of the confined
  fluid is fixed at $\rho=(6/\pi)0.4$. We compare molecular dynamics
  simulation data
  (\opencircle) for the confined fluid with the self diffusivity of a 
  bulk HS fluid, $\dbx$, at the same value of $x=\rho$ (green
  dash-dotted line), 
  $x=\exs$ (red square), and $x=\bpo$ (blue triangle). Solid lines are
  shown as a guide to the eye. $D$ and $\exs$, calculated via molecular dynamics and TMMC
  simulations, respectively, are taken 
from Fig. 7 of~\cite{Mittal2007Doesconfininghard-sphere}. $\bpo$ was
  calculated via TMMC simulations described in Section~\ref{sec:methods} 
of the present work.
  \label{D_osscilations}}
\end{figure}
Another relevant test case for comparing which of $\rho$, $\exs$, or 
$\bpo$ most faithfully tracks dynamics
is to vary the degree of confinement while fixing $\rho$, an idea
motivated by an earlier study by Mittal
\etal~\cite{Mittal2007Doesconfininghard-sphere}.  In particular,
Mittal \etal demonstrated that $D$ and $\exs$ of a HS fluid 
oscillate in phase when $H$ of the confining slit pore is varied (for $H \le
5$) and $\rho$ is held constant.  
The maxima in $D$ (high particle mobility)
and $\exs$ (weak interparticle correlations) occur 
for integer values of $H$, geometries which naturally accommodate 
the layering of particles near the boundaries. The minima in 
$D$ (low particle mobility) and $\exs$
(strong interparticle correlations) occur for non-integer values of
$H$, which frustrate this natural layering pattern.  Along similar
lines, Goel
\etal~\cite{Goel2008Tuningdensityprofiles} recently demonstrated that
particle-boundary interactions that flatten the density profile of a
confined fluid generally reduce $D$ and $\exs$, while those which
increase layering can have the opposite effect.   
All of this suggests that excess entropy captures some of the subtle
frustration induced effects that layering has on both interparticle
correlations and single-particle 
dynamics~\cite{Mittal2007Doesconfininghard-sphere}.  Does $\bpo$ also capture
these effects?  A very recent study of Mittal
\etal~\cite{Mittal2008LayeringandPosition} suggests that it might.  In 
particular, the authors of that study showed that the local fraction of
available volume $p_0(z)$ and the position-dependent self diffusivity
normal to the boundaries $D_{\perp}(z)$ of a confined HS fluid 
are highest in regions of high local density $\rho(z)$. 

Figure~\ref{D_osscilations} provides a more direct test of this idea.
In particular, it shows the $D$ data of 
Mittal~\etal~\cite{Mittal2007Doesconfininghard-sphere} for a HS fluid
confined between hard walls calculated via
molecular dynamics simulations.  We have also included on this plot 
predictions from the three bulk structure-property relations $\dbx$,
where $x=\rho$, $x=\exs$, and $x=\bpo$.  Since $\rho$ is fixed here, 
it is evident that $D_{\mathrm{bulk}}^{\rho}$ is not able to
predict the oscillatory trends in the dynamics data.
However, note that 
both $D_{\mathrm{bulk}}^{\exs}$ and $D_{\mathrm{bulk}}^{\bpo}$ 
predict the correct oscillatory behavior.  In fact, the predictions of 
$D_{\mathrm{bulk}}^{\bpo}$ are virtually quantitative over the entire
range of $H$.  

\begin{figure}[h]
  \includegraphics{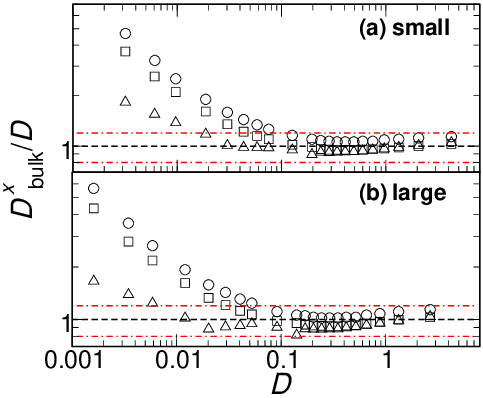}
  \caption{Ratio of self diffusivity of (a) small particles and (b)
    large particles of a bulk binary HS fluid mixture to
    that of a corresponding confined HS fluid mixture with the same 
    value of a static quantity $x$, $\dbx/D$, obtained 
    via molecular simulation. The static quantity used for making
    predictions is $x=\rho$ (\opencircle), $x=\exs$ (\opensquare), 
    and $x=\bpo _i$ (\opentriangle). 
    Data are shown for the $H=5$ slit-pore
    geometry with the total packing fraction in the range
    $0.025-0.52$.  The
    mole fraction of the small spheres is $0.75$. $20\%$ bounds on
    relative error in self-diffusivity predictions
    are shown by red dash-dotted line.\label{dbydbulk_slitmix}}
\end{figure}
Should we expect $D_{\mathrm{bulk}}^{\bpo}$ to generally 
track the dynamics of dense, confined HS fluids
more accurately than $D_{\mathrm{bulk}}^{\exs}$?  
In other words, what is more relevant for dynamics of inhomogeneous fluids: 
available space or available states?  We further explore that question
by examining the behavior of the confined HS
mixture discussed in Section~\ref{sec:methods}.  Specifically, by
studying this binary fluid mixture in a slit pore with $H=5$, we
are able to probe confined fluid states with packing fraction $\phi$
as high as 0.52 (compared with the highest 
packing fraction of $0.46$ for a monodisperse HS fluid confined 
in a slit-pore of $H=5$) At $\phi=0.52$, the fluid already exhibits 
dynamic signatures of supercooling, e.g., the emergence of a
plateau in the time dependence of the mean-squared displacement. 
The corresponding $D$ at this packing fraction 
($=0.002$) is an order of magnitude smaller than the smallest
$D$ for the confined monodisperse fluid ($=0.02$) in the slit pore 
geometry. 

Figure~\ref{dbydbulk_slitmix}(a,b) shows the ratio of the bulk
self diffusivity to confined self diffusivity, $\dbx/D$, for the small
and large particles of the mixture, respectively.  Again, the comparisons
are made to the bulk fluid mixture of the same composition and 
density ($x=\rho$), excess entropy
($x=\exs$), or fractional available volume of the
corresponding species ($x=\bpo_i$).  Note that the 
bulk structure-property predictions for self diffusivities of
small and large particles are semi-quantitative (within
$20\%$) for $D>0.1$ based on any of the three
aforementioned static quantities.  However, relative 
errors in predictions based on $\rho$ or $\exs$ begin to increase 
sharply for $D<0.1$.  On the other hand, predictions based on $\bpo_i$
remain semi-quantitative for all $D>0.02$ (covering three decades in
$D$), with significant
overpredictions occuring only for the densest three state points 
investigated.  Thus, it appears that, for single-particle dynamics, 
fractional available volume is
the most relevant of the three static measures investigated here.
The question of whether there exists an alternative static measure
$x$, such that $\dbx$ tracks $D$ for deeply supercooled mixtures,
is currently an open one.  The answer to that question will likely have
important implications for understanding how confinement shifts the glass
transition of fluids. 
       
\subsection{Particle-boundary interactions and a generalized measure
  of available volume}
\label{sec:dbydbulk_neutral}

Thus far, we have only considered the geometric (i.e., packing)
consequences of confinement on dynamics.  
How does the presence of finite particle-boundary interactions 
affect the picture described in the previous section? 
We explore the answer to this
question by studying a monodisperse HS fluid confined to a slit pore
geometry by smooth walls with either square-shoulder (repulsive) or
square-well (attractive) particle-boundary interactions 
[for details, see Eq.~\eref{sw_interaction}].  

\begin{figure}[h]
  \includegraphics{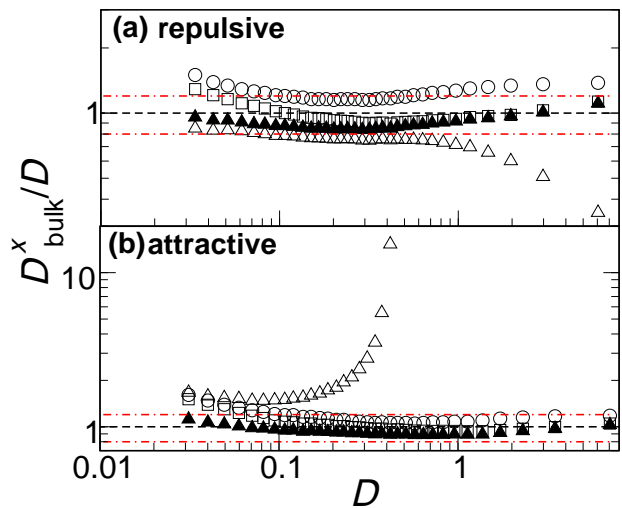}
  \caption{Ratio of self diffusivity of a bulk HS fluid to
    that of a confined HS fluid with the same 
    value of a static quantity $x$, $\dbx/D$, obtained 
    via molecular simulation. Data include systems with (a) square-shoulder (repulsive) and (b) 
square-well (attractive) particle-boundary interactions (see
Eq.~\ref{sw_interaction}) with 
$H=5$ in the slit pore geometry. 
The static quantity used for making predictions is $x=\rho$
(\opencircle), $x=\exs$ (\opensquare), $x=\bpo$ (\opentriangle), and
$x=\bpow$ (\fulltriangle). $20\%$ bounds on relative error in self-diffusivity predictions
    are shown by the red dash-dotted line.
Particle center accessible density for the fluid spans the range $0<\rh<1.05$.
     $\bpo$ overpredicts high-diffusivity state points in panel
     \textit{b} by more than $1000\%$ and those data points are off
     the scale of the graph. \label{dbydbulk_slitint}}
\end{figure}

Figure~\ref{dbydbulk_slitint}(a and b) 
shows the ratio of the bulk to confined self diffusivity, $\dbx/D$,
for these two cases, respectively,
with comparisons between bulk and confined fluids being 
made at the same density ($x=\rho$), excess entropy
($x=\exs$), and fractional available volume ($x=\bpo$). Clearly, the fractional available volume $\bpo$ fails to track 
the dynamics of the confined fluid for
both cases presented.  However, this should perhaps be expected.
Since the interactions of the particles with the boundaries in these
cases are strongly position-dependent, all free space is not 
equally ``available'' to the particles.  To account for this 
energetic imbalance, we suggest a generalized
available volume ($\bpow$) that appropriately weighs the local available
space by the Boltzmann factor of the particle-boundary interaction,
\begin{eqnarray}
  \bpow & = & \frac{\int_{V_\mathrm{c}} \po \exp [-u_\mathrm{w} (s)]dV}{\int_{V_\mathrm{c}}\exp [- u_\mathrm{w}(s)] dV}
  = \frac{\rh (\xi)}{\rho_\mathrm{h}^\mathrm{ig} (\xi)} \nonumber \\
  & = & \frac{\xi ^\mathrm{ig}(\rh)}{\xi (\rh)}=\exp[-\{\mu(\rh)-\mu^\mathrm{ig}(\rh)\}] 
\label{eq:GFV}
\end{eqnarray}
where, in all cases, the superscript ``ig'' denotes the corresponding
quantity for an ideal gas confined to an identical slit pore.  As one
can see from the above equation, since this generalized available volume 
inherently relates the thermodynamic state of the confined fluid to
that of an ideal gas, it bears some resemblence to an excess
thermodynamic property.

It is important to point out that this generalized available volume 
has several distinguishing
features. (i)~It reduces to $\bpo$ in the limit of fluids confined
to smooth hard boundaries.  Thus, all of the results presented earlier 
in this paper for $\bpo$ will remain unchanged for those systems if
one instead uses $\bpow$.  (ii)~Unlike $\bpo$ or density, there is no
arbitrary choice that needs to be made about the volume over which
one should do the averaging.  This quantity is the same no matter
whether averaging is carried out over the center-accessible or the 
total volume of the fluid.  Put differently, this definition removes 
any arbitrariness about the effective ``diameter'' of the 
fluid-boundary interaction.  
(iii)~The quantity $\bpow$ can be computed directly 
from knowledge of standard thermodynamic and system parameters, 
namely $\rho$, $\xi$, and $u_\mathrm{w} (z)$.  (iv) It is not limited
to HS fluid systems.  In fact, the computation of $\bpow$ from Eq.~\eref
{eq:GFV} does not even require information about 
the particle-particle interactions, as long as the other thermodynamic 
quantities can be measured.

How well does this new generalized measure of available volume track 
self diffusivity when finite particle-boundary interactions are
present?  Figure~\ref{dbydbulk_slitint} clearly illustrates that 
$\bpow$ corrects for the problems that $\bpo$ faces in predicting
the dynamics in these cases.   In particular, the maximum error 
in self-diffusivity predictions based on $\bpow$ 
is $16\%$ across the entire range of packing fractions investigated,
which makes it a more reliable predictor of single-particle dynamics
than either $\exs$ or $\rho$ for these systems.  As was seen earlier, 
diffusivity predictions based on the bulk structure-property relation for
$\bpow$ are considerably more accurate than that for $\rho$ or $\exs$ 
when considering high density, low diffusivity state points ($D<0.1$).  

\section{Using DFT together with bulk structure-property
  relations to predict dynamics of confined fluids}
\label{sec:dbydbulk_DFT}

\begin{figure}[h]
   \includegraphics{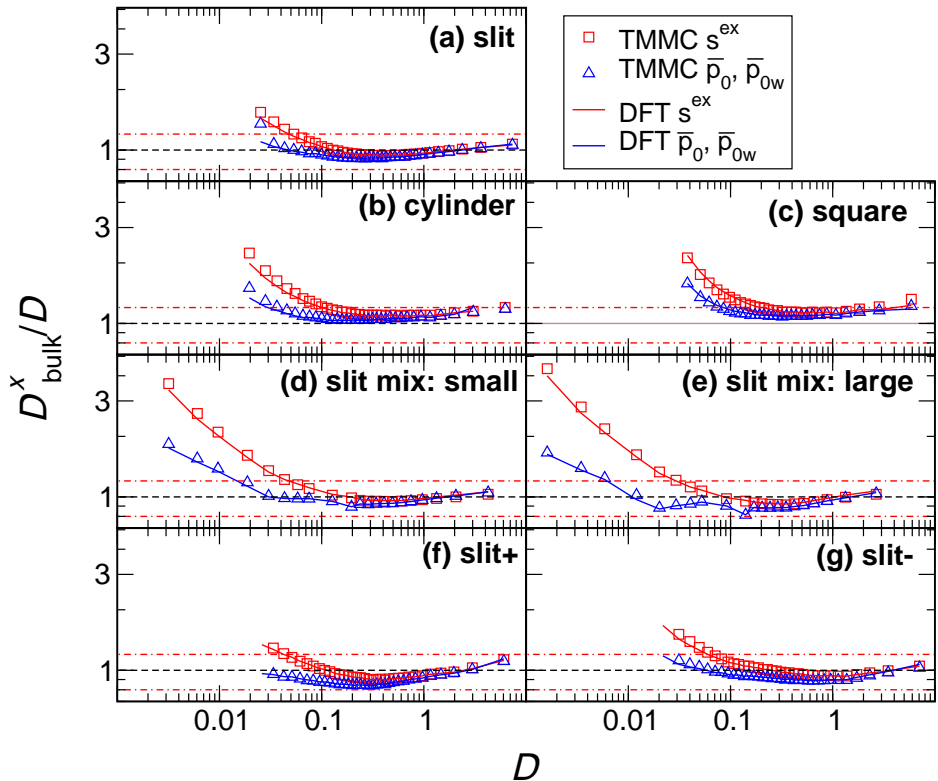}
  \caption{Ratio of self diffusivity of a bulk HS fluid to
    that of a confined HS fluid with the same 
    value of a static quantity $x$, $\dbx/D$. Here we show a
    comparison of $\dbx/D$ with $x$ obtained from the DFT (lines) or 
    TMMC (symbols) calculations. (a, b, and c) Data for a 
    monodisperse HS fluid confined by smooth hard 
    boundaries to a slit pore [$H=5$], a cylindrical pore [$H=6$], 
    and a square channel [$H=5$] geometry, respectively. 
    (d and e) Data for small and large
    particles of a binary HS mixture confined in a $H=5$ slit
    pore, respectively. (f and g) Data for a monodisperse HS fluid
    confined between the repulsive and attractive boundaries discussed
    in the text, respectively, 
    placed at a separation of $H=5$ in slit pore geometry. 
    The density range for the state points shown are the same as in
    Figure~\ref{dbydbulk_all} for [a-c], 
    Figure~\ref{dbydbulk_slitmix} for [d and e], and
    Figure~\ref{dbydbulk_slitint} for [f and g].
    $20\%$ bounds on relative error in self-diffusivity predictions
    are shown by the red dash-dotted line.\label{dbydbulk_DFT}}
\end{figure}

Above we have shown that knowledge of $\dbx$ for the bulk HS fluid
($x=\exs$ or $x=\bpow$) together with the value of $x$ in confinement
is enough for semi-quantitative prediction of confined (equilibrium) fluid
diffusivity, $D$, across a wide range of parameter space in these systems. 
Thus far, we have used TMMC simulations to determine $x$ for each confined
fluid of interest.  That raises the following question.  Is classical 
DFT accurate enough in its
predictions of $x$ that one can eliminate the step of simulating the
confined fluid altogether?  In this section, we take a first step toward addressing this question.  
In particular, we 
present calculations of the ratio of self diffusivity of a bulk HS
fluid to that of a confined HS fluid with the same value of $x$,
$\dbx/D$, where $x=\exs$ or $\bpow$.  In each case, $x$ 
of the confined fluid is obtained directly from predictions of
Rosenfeld's fundamental measure
theory~\cite{Yu2002Structuresofhard-sphere}, an accurate DFT for these systems.

Figure~\ref{dbydbulk_DFT} shows relative errors 
in self-diffusivity predictions based 
on $\exs$ and
$\bpow$, comparing cases with knowledge of the ``exact'' value of $x$
in confinement  (calculated
via TMMC simulations) and the predicted value of $x$ (calculated via DFT).
Selected cases explored in the previous section 
involving the three confining geometries (slit pore,
cylindrical pore, and square channel), small and large particles of 
binary mixtures, and finite
fluid-boundary interactions are presented.  In all cases, the
$\dbx/D$ curves obtained via the two routes (TMMC versus DFT) 
are virtually indistinguishable over the entire density range of
equilbrium fluid, a demonstration of the reliability of
DFT for computing the static properties of these systems. 
As a result, it is clear that one can use the bulk structure-property
relations discussed above together with predictions of $x$ from DFT to make 
semi-quantitative estimates of confined fluid self diffusivity for a
wide variety of hard sphere systems.   

\section{Conclusions}
\label{sec:theend}

Fluids trapped in small spaces feature prominently in science and 
technology, and understanding their properties is key for a number of 
research areas that range from the design of membranes 
to the engineering of microfluidic devices.  
The static and dynamic properties of these confined fluids 
can be very different than those of bulk samples.  
While quantitatively accurate theories like DFT are available for 
predicting static properties of confined fluids, making even 
qualitative predictions for dynamics of inhomogeneous fluids 
has long been a challenging endeavor.  
In this paper, we demonstrate how semi-quantitative (albeit indirect) 
predictions of
dynamics from first principles are still possible, even in the absence 
of a theory, once one recognizes that certain relationships between
static and dynamic properties are insensitive to confinement.  

This study provides a systematic and quantitative 
investigation of such relationships.
In particular, we present a comprehensive
study of the effects of confinement on the correlation between
self diffusivity ($D$) and various thermodynamic measures for confined HS
fluids:  particle
center-accessible-volume-based and total-volume-based 
average densities ($\rh$ and $\rho$, respectively), excess
entropy ($\exs$), and two average measures of fractional available volume 
($\bpo$~and~$\bpow$). Our main findings are as follows.  The 
bulk structure-property correlation, $\dbx$,  
based on the first 
density measure, $x=\rh$, severely underestimates $D$ when $\rh$ of
the confined fluid is used as the input.  Further, for 
dense confined fluids, $\rh$ is often larger than the freezing
density of the bulk fluid, eliminating altogether 
the possibility of using the corresponding 
bulk structure-property relation for predictions.  Self-diffusivity
predictions based on the relation with 
total-volume-based density $x=\rho$ provide a
significant improvement over those involving $\rh$, substantiating the
earlier idea that $\rho$ might be
considered as a more natural measure of density for predicting
dynamics~\cite{Mittal2006ThermodynamicsPredictsHow}. 

However, when one considers a wider 
variety of geometries [slit pore, cylindrical,
square channel], confined fluid mixtures, finite 
particle-boundary interactions, and a wide range of packing fractions,
one finds that 
structure-property relations $\dbx$ based on excess entropy ($x=\exs$)
and a new generalized measure of available volume ($x=\bpow$) 
provide much more accurate
estimates for $D$ than those based on either of $\rh$ or $\rho$. 
Further, self-diffusivity predictions based on $\bpow$ are 
significantly more accurate than those based on $\exs$ under
conditions of high packing
fractions (e.g., supercooled fluids) and highly restrictive
confining geometries (e.g., quasi-one-dimensional channels).  The generalized
available volume $\bpow$ may also be easier to compute based on
experimental quantities than $\exs$,
since the former is related to average density and chemical potential in a
simple way (see Eq.~\ref{eq:GFV}).  Importantly, neither $\bpow$ nor
$\exs$ require one to arbitrarily define an averaging volume for the 
confined system.  

Predictions of $x$ via classical DFT are accurate
enough for inhomogeneous HS fluids that one can use them, together
with the bulk structure-property relation ($\dbx$), to make
semi-quantitative estimates of confined fluid diffusivities.
This effectively eliminates the need for simulating the confined fluid 
altogether, which might be particularly convenient in applications 
where one needs to estimate the dynamics of systems
across a wide array of parameter space.  For example, in 
the design of microfluidic 
systems, one might hope to screen a large range of 
possible particle-boundary interactions or confining geometries 
against design considerations.  A
preliminary application of this idea~\cite{Goel2008Tuningdensityprofiles} is to use DFT to passively tune the transport properties of a confined fluid, in a
controlled way, by modifying the geometry or boundary-particle
interactions of the confined space.

Can the
aforementioned static measures predict the dynamics of fluids with
continuous intermolecular potentials and/or attractive interactions?
As mentioned earlier, recent data from molecular simulations have
shown that there is indeed an 
isothermal correlation between the self-diffusion
coefficient $D$ and the excess entropy for a variety of confined
fluids (e.g., Lennard-Jones, square-well, Weeks-Chandler-Andersen),
approximately independent of the degree of
confinement for a wide range of equilibrium
conditions~\cite{Mittal2007RelationshipsbetweenSelf-Diffusivity,Goel2008Tuningdensityprofiles}.
We are currently exploring the viability of using the generalized
available volume for predicting dynamics in these fluids, and we will
report on our findings in a future study.




\ack
T.M.T. acknowledges support of the National
Science Foundation (NSF) under Grant No. CTS-0448721, the Welch
Foundation, the David and
Lucile Packard Foundation, and the Alfred P. Sloan
Foundation. W.P.K. and G.G. acknowledges support from a NSF Graduate Research
Fellowship and a UT ChE department fellowship, respectively. The Texas Advanced
Computing Center (TACC), the Biowulf PC/Linux cluster at the 
National Institutes of Health, Bethesda, MD, and the University at
Buffalo Center for Computational Research provided computational 
resources for this study.


\end{document}